\documentclass[pra,twocolumn,showpacs,amsmath,amssymb]{revtex4}
\usepackage{graphicx}
\usepackage{dcolumn}
\usepackage{bm}
\begin{document}
\preprint{}
\title{Noise and instability of an optical lattice clock}
\author{Ali Al-Masoudi, S\"oren D\"orscher, Sebastian H\"afner, Uwe Sterr, Christian Lisdat}
\affiliation{Physikalisch-Technische Bundesanstalt, Bundesallee 100, 38116 Braunschweig, Germany}
\date{October 23, 2015} 
\begin{abstract}
We present an analysis of the different types of noise from the detection and interrogation laser in our strontium lattice clock. We develop a noise model showing that in our setup quantum projection noise--limited detection is possible if more than 130~atoms are interrogated. Adding information about the noise spectrum of our clock laser with sub-$10^{-16}$ fractional frequency instability allows to infer the clock stability for different modes of operation. Excellent agreement with experimental observations for the instability of the difference between two interleaved stabilizations is found. We infer a clock instability of $1.6 \times 10^{-16}/\sqrt{\tau / \mathrm{s}}$ as a function of averaging time $\tau$ for normal clock operation.
\end{abstract}

\maketitle
%
%
%
%
%
%
\section{\label{sec:intro}Introduction}
Optical clocks in general and lattice clocks in particular have shown outstanding fractional frequency instabilities of about $2 \times 10^{-16}/\sqrt{\tau / \mathrm{s}}$ \cite{nic15,hin13,nic12}, averaging down to the low $10^{-18}$-regime. This development has been made possible mainly by significant improvements on the frequency stability of the interrogation lasers \cite{jia11, nic12, hag13, hae15a}. The improvement of clock stability has a strong impact on the determination of the clock's accuracy since systematic frequency shifts can be evaluated with higher accuracy in reasonable time. Furthermore, the stability of a clock  also determines how practical it is for actual measurements and if, e.g., temporal variations of signals \cite{der14} can be observed.

The instability of a clock stems from different sources of noise contributing to the error signal detected in the clock cycle, i.e.\ the estimated frequency offset of the clock laser from the atomic transition, and from the Dick effect \cite{dic87} due to the non-continuous interrogation of the atoms by the clock laser \cite{wes10a}. The Dick effect originates mostly from the unobserved clock laser fluctuations during the dead-time of the clock cycle. 

In high-performance clocks the reference transition is coherently interrogated by either Rabi or Ramsey schemes on a frequency-sensitive slope of the spectroscopic signal. In the detection, the quantum superposition state of each individual atom is projected to the ground or excited state, leading to the fundamental noise limit given by the quantum projection noise (QPN) \cite{ita93}. An increase of atom number $N$ will reduce the influence of the projection noise as it scales with $\sqrt{N}$ while the signal scales with $N$. However, in practice the atom number is limited by other factors as, e.g., collision shifts \cite{rey14, lem11, lud11, gib09}. Ideally, the QPN should be the dominant noise contribution in the measurement of the frequency offset of the local oscillator from the atomic line. 

Squeezed states \cite{win92} and entanglement of the atoms \cite{kes14} can overcome the projection noise limit. It must, however, be noted that at present even the most stable optical clocks are still limited by the aliasing of laser noise, i.e., via the Dick effect. 

With this background it becomes obvious that a detailed understanding of the noise sources present in the experiment is not only essential for optimizing the clock stability and interrogation strategy, but also for judging the necessity to implement squeezing and entanglement methods to improve the clock. 

We therefore give a rigorous analysis of the detection noise contributions in our Sr lattice clock \cite{fal14} (Sec.~\ref{sec:noise}). The noise model is further supplemented by a contribution of the clock laser noise \cite{hae15a} via the Dick effect (Sec.~\ref{sec:laser}). Combining both, we show that the clock instability observed during evaluations of systematic effects in interleaved stabilization mode is well reproduced and a clock instability of $1.6 \times 10^{-16}/\sqrt{\tau / \mathrm{s}}$ can be inferred for regular clock operation (Sec.~\ref{sec:stability}), which is governed by the excellent stability of our clock laser \cite{hae15a} and competitive with the best values of $2.2 \times 10^{-16}/\sqrt{\tau / \mathrm{s}}$ \cite{nic15} and $3.2 \times 10^{-16}/\sqrt{\tau / \mathrm{s}}$ \cite{hin13} published to date. 
%
%
%
%
%
%
\section{\label{sec:noise}Detection noise of the lattice clock}
In this section, the individual noise contributions affecting the measured excitation probability are discussed. In the final part of the section, their influence on the clock signal is modeled and compared with experimental observations. For the sake of simplicity, we express all noise amplitudes in arbitrary units of the data acquisition system labeled `counts'.

\begin{figure}
\centerline{\includegraphics[width=8 cm]{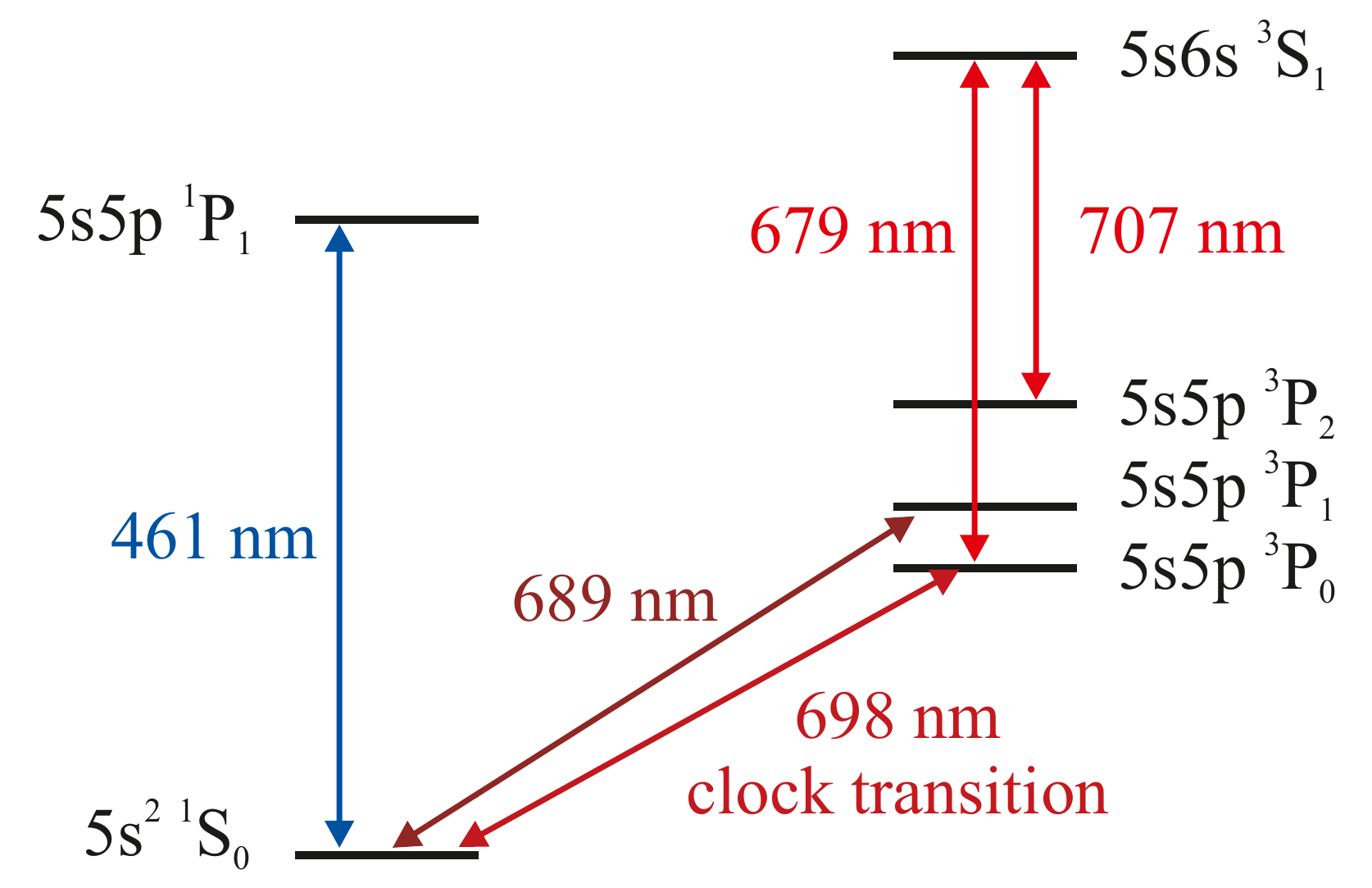}}
\caption{Simplified level scheme of the strontium atom. Arrows indicate the transitions used for cooling and spectroscopy and their associated wavelengths.}
\label{fig:scheme}
\end{figure}

Figure~\ref{fig:scheme} shows a simplified level scheme of the strontium atom. Details for laser cooling and trapping of strontium atoms have been described in previous publications \cite{fal14, fal11, lis09}. To derive the spectroscopic signal of the lattice clock after the interrogation of the  $^{87}$Sr atoms by the clock laser, atoms in the ground state (${^1S_0}$) are excited with a resonant laser beam in standing-wave configuration on the strong 461~nm cooling transition, ${^1S_0} - {^1P_1}$. The fluorescence is observed by a photomultiplier tube in current mode. Its signal is amplified and digitized with an analog-to-digital converter of the data acquisition computer. This signal $g$ is, apart from an offset $o$, proportional to the ${^1S_0}$ ground state atom number. The radiation pressure removes the atoms from the detection volume within 20~ms.
	
Similarly, the signal $e$ from atoms in the excited clock state (${^3P_0}$) is detected after the atoms have been optically pumped within few 100~$\mu$s to the ${^3P_1}$ state by two lasers resonant with the ${^3P_0} - {^3S_1}$ and ${^3P_2} - {^3S_1}$ transitions, from where they decay further to the ground state. In addition, the offset $o$ in both $g$ and $e$ due to stray light, multiplier dark current, and electronic offset is measured via a third and final detection pulse after removing any remaining atoms. Therefore, each of the signals $g$, $e$, and $o$ is a sum of different contributions from fluorescence ($S_\mathrm{fluo}$), stray light ($S_\mathrm{stray}$), and electronic background ($S_\mathrm{elec}$). From these three signals, the atomic excitation probability $p_\mathrm{e}$ is estimated by 
\begin{equation}
p_\mathrm{e} = \frac{e - \overline{o}}{e + g -2\overline{o}}.
\label{eq:pe}
\end{equation}
This normalization removes substantial noise on $g$ and $e$ due to shot-to-shot fluctuations of the atom number $N$. Here, we replaced the measured offset $o$ by its running average $\overline{o}$, since the variation of the actual background is small on timescales of several interrogation cycles; thus the additional noise of $p_\mathrm{e}$ due to the shot-to-shot noise of the offset measurement is suppressed and does not need to be considered in the noise model.

The connection between the excitation probability, in particular its noise, and the corresponding frequency excursion of the interrogation laser is given by the slope of the spectroscopic signal. For the case of probing the atomic resonance line at a half-width point \cite{dic87} using Rabi interrogation with a pulse of length $T_\mathrm{\pi}$, the slope is
\begin{equation}
\frac{\mathrm{d}p_\mathrm{e}}{\mathrm{d}\nu} \approx \pm 2 \pi \cdot 0.30 \cdot T_\mathrm{\pi}
\label{eq:slope}
\end{equation}
with its sign depending on which side of the resonance is probed. A Ramsey interrogation scheme with a free precession time $T_\mathrm{Ramsey}$ and short excitation pulses leads to a steeper slope of
\begin{equation}
\frac{\mathrm{d}p_\mathrm{e}}{\mathrm{d}\nu} \approx \pm 2 \pi \cdot 0.5 \cdot T_\mathrm{Ramsey}.
\label{eq:slopeRam}
\end{equation}
%
%
\subsection{\label{sub:electronic}Electronic noise}
%
The electronic noise was measured by running the standard detection sequence without atoms or laser light being present. The observed noise is a sum of amplifier, digitizer, and dark current noise. Such noise will be present in the three signals, $g$, $e$, and $o$, but not in $\overline{o}$. The standard deviation of the signal is $\sigma_\mathrm{elec} = 0.82$~counts, while the amplitude of the electronic offset signal $S_\mathrm{elec} \approx 18$~counts.
%
%
\subsection{\label{sub:shot}Photon shot noise}
\begin{figure}
\centerline{\includegraphics[width=8 cm]{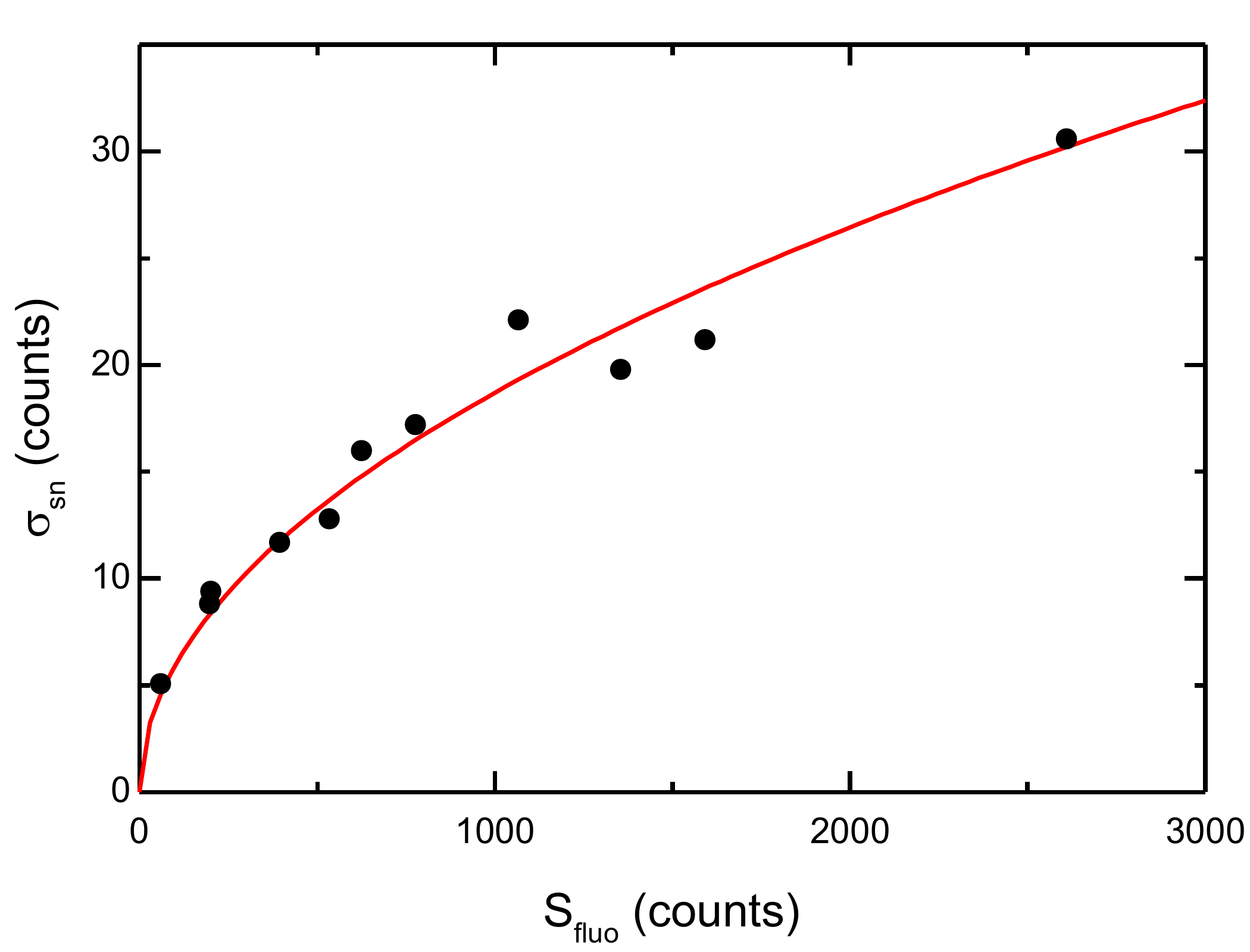}}
\caption{Measured fluorescence shot noise levels $\sigma_\mathrm{sn}$ (full circles) as a function of the detected signal $S_\mathrm{fluo}$ along with a fit according to Eq.~\ref{eq:sn} (solid line).}
\label{fig:shotnoise}
\end{figure}

%
During the detection of the atomic fluorescence, a finite number of photons is collected by the photomultiplier tube. Thus, the signals $g$ and $e$ will suffer from a shot noise contribution. In order to quantify this contribution, we have investigated the photon shot noise using a flashlight as a shot noise--limited light source. For the observed white phase noise, the first point of the Allan deviation of the recorded data is equal to their standard deviation; we use this value as a measure of the noise to remove the influence of slow intensity variations of the flashlight. The measurements were performed for different signal amplitudes. The electronic background $S_\mathrm{elec}$ (discussed in the previous subsection) must be removed to extract the actual `fluorescence' signal  $S_\mathrm{fluo}$, i.e.\ the signal stemming only from detected photons. The resulting  `fluorescence' noise contribution $\sigma_\mathrm{sn}$ is shown in Fig.~\ref{fig:shotnoise}. The expected $\sqrt{S_\mathrm{fluo}}$-dependence of the noise amplitude $\sigma_\mathrm{sn}$ is well reproduced. From a fit we find 
\begin{equation}
\sigma_\mathrm{sn} (S_\mathrm{fluo}) = 0.59(2) \sqrt{S_\mathrm{fluo}},
\label{eq:sn}
\end{equation}
which corresponds to $S_\mathrm{fluo}$ = 0.35(3) counts per detected photon. Fluorescence shot noise will be present in the signals $g$ and $e$ with an amplitude depending on the fluorescence contribution $S_\mathrm{fluo}$ to these signals, as e.g.\ the electronic offset $S_\mathrm{elec}$ is not subject to shot noise.
%
%
\subsection{\label{sub:laser}Detection laser intensity noise}
%
Intensity fluctuations of the detection laser at the position of the atoms will show up on the fluorescence signal as long as the transition is not strongly saturated. We avoid high intensity, since the radiation pressure--induced heating of the atoms reduces the reasonable interaction time and thus the detected signal. This loss of signal is not compensated by the higher photon scattering rate, and the detection would be less efficient. 

Intensity fluctuations may arise from both power and pointing instability of the detection laser. We measured the power fluctuations of the laser in a similar procedure as for the shot noise measurements, except that we used stray light from the detection laser instead of a flashlight. From these measurements we have found data that is very similar to the one presented in Fig.~\ref{fig:shotnoise}. In particular, we have observed no significant noise contribution with a linear dependence on laser power and thus conclude that laser power noise on short time scales, which would result in such a contribution, can be neglected. This is corroborated further by direct measurements of the laser power. The optical setup of the detection beam, which is delivered by a fiber and collimated to a diameter of about 2~mm, comprises only a short free-space path on the order of 50~cm; we have analyzed its pointing instability and found it to be negligible on relevant time scales. We also note that the shot noise of the detection beam with a power of about 400~$\mu$W is negligible.

Laser stray light ($S_\mathrm{stray}$) adds a shot-noise contribution to the signals $g$ and $e$ according to Eq.~\ref{eq:sn}, whereas the similar noise contribution to the offset is suppressed by the use of $\overline{o}$.
%
%
\subsection{\label{sub:freq}Detection laser frequency noise}
%
The 461~nm detection laser beam is derived from a frequency-doubled diode laser system that also produces the laser beams for laser cooling and Zeeman slowing. The fundamental frequency of the laser system is stabilized to a high-finesse, 10~cm long ultra-low expansion glass (ULE) resonator. With its resonance line width of less than 100~kHz, we estimate in-lock frequency fluctuations on the kilohertz level from the in-loop error signal. Compared to the 32~MHz line width of the ${^1S_0} - {^1P_1}$ detection transition, frequency noise of the detection laser does not contribute significantly to the  detection noise.
%
%
\subsection{\label{sub:QPN}Quantum projection noise}
%
%
\begin{figure}
\centerline{\includegraphics[width=8 cm]{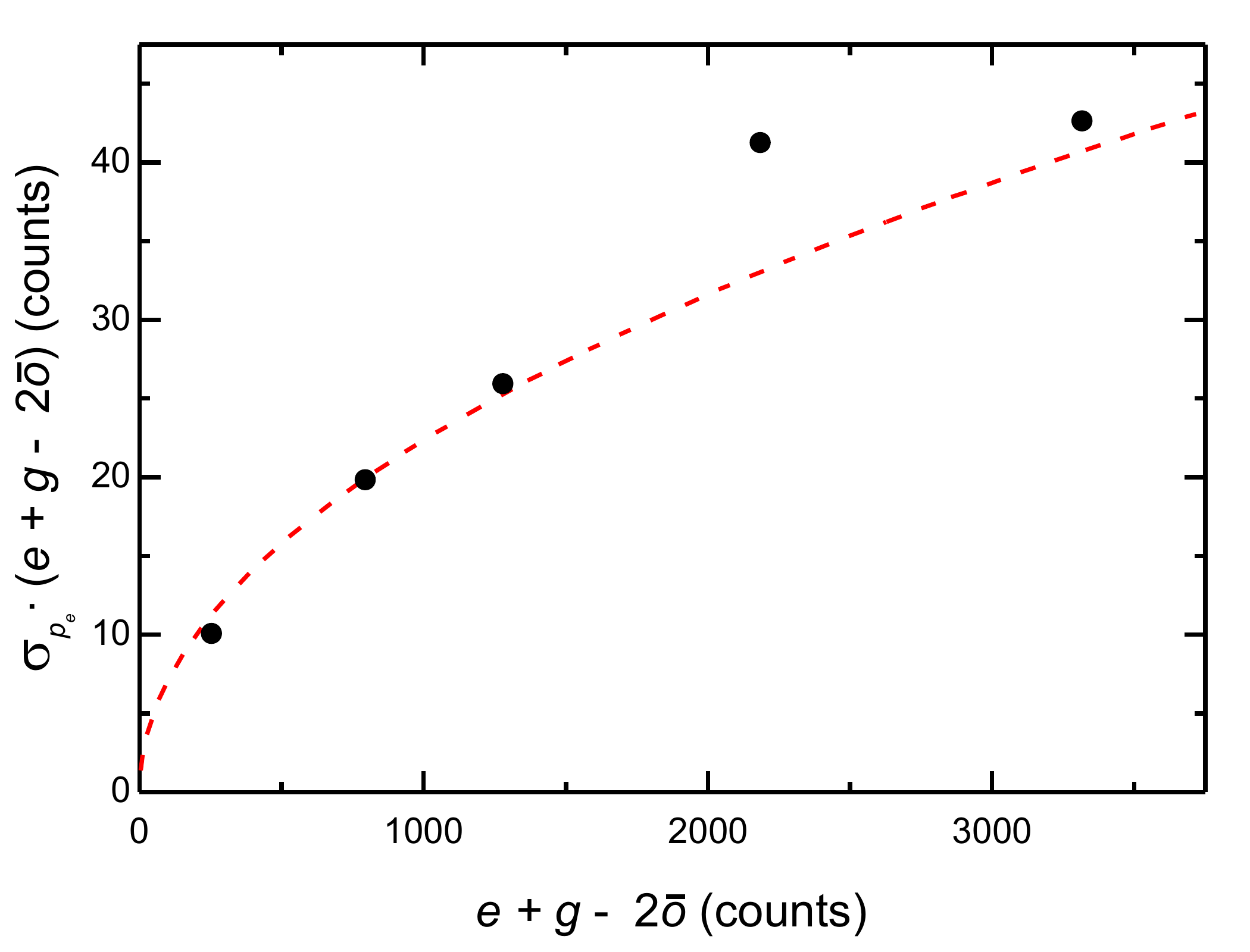}}
\caption{Excitation probability noise $\sigma_{p_\mathrm{e}}$ at $p_\mathrm{e} = 0.5$ multiplied by the signal amplitude $g + e -2\overline{o}$ as a function thereof (full circles). Atoms are prepared in a superposition of ${^1S_0}$ and ${^3P_0}$ states by a frequency-insensitive resonant $\pi/2$-pulse. The dashed line shows a fit of Eq.~\ref{eq:QPN} with a factor of proportionality close to the experimentally expected one (see text).}
\label{fig:QPN}
\end{figure}
As mentioned in Sec.~\ref{sec:intro}, QPN is the fundamental noise that ideally should dominate all other noise contributions in the detection. It depends on the atom number $N$ and the excitation probability $p_\mathrm{e}$  \cite{ita93} as
\begin{equation}
\sigma_\mathrm{QPN} (p_\mathrm{e}) = \sqrt{p_\mathrm{e} (1 - p_\mathrm{e}) N}.
\label{eq:QPN}
\end{equation}
The signal-to-noise ratio $N/\sigma_\mathrm{QPN}$ can thus be improved by increasing the atom number. $N$ is proportional to $g + e -2\overline{o}$, where the constant of proportionality to convert from `counts' to atom number is about one atom per count. This factor depends on the actual alignment of the experiment, the frequency and power of the detection laser, and other parameters.

We measured the noise $\sigma_{p_\mathrm{e}}$ of the excitation probability of  samples of atoms with $p_\mathrm{e} = 0.5$ and different atom numbers $N$ to investigate whether we can achieve QPN-limited detection in our experiment.  In order to become insusceptible to frequency noise of the interrogation laser, we prepared a coherent superposition state by using a resonant $\pi/2$-Rabi pulse instead of Rabi excitation with a $\pi$-pulse at a half-width frequency detuning as usually applied in a stabilization sequence. A pulse length of $T_\mathrm{\pi/2} = 10.5$~ms was chosen. The observed noise $\sigma_{p_\mathrm{e}}$ is plotted in Fig.~\ref{fig:QPN}. To which extent quantum projection noise is dominant in our setup and if residual laser noise affected the data will only become apparent with the combined noise analysis in Sec.~\ref{sub:det}.
%
%
\subsection{\label{sub:det}Detection noise model}
%
Having quantified the individual noise sources, we now develop a noise model to combine them and verify if the observed detection noise (Fig.~\ref{fig:QPN}) is fully described and whether QPN is the dominant noise source.

The noise model assumes independent contributions to $\sigma_{p_\mathrm{e}}$ from QPN and other sources of noise.
QPN is handled separately because it leads to anti-correlated noise in $g$ and $e$. For the other noise sources, independent contributions from $g$ and $e$, but not from $\overline{o}$, are considered. For these contributions, the model uses the derivatives of $p_\mathrm{e}$ (Eq.~\ref{eq:pe}) with respect to $g$ and $e$. Thus the total noise is given by
\begin{equation}
\sigma_{p_\mathrm{e}} = \sqrt{\sum_{i} \left( \frac{\mathrm{d}p_\mathrm{e}}{\mathrm{d}g} \sigma_{g,i} \right)^2 + \sum_{i}\left( \frac{\mathrm{d}p_\mathrm{e}}{\mathrm{d}e} \sigma_{e,i} \right)^2 + \frac{\sigma^2_\mathrm{QPN}}{N^2}}
\label{eq:noisemod}
\end{equation}
with the derivatives
\begin{eqnarray}
 \frac{\mathrm{d}p_\mathrm{e}}{\mathrm{d}g} &=& \frac{\overline{o}-e}{(g+e-2\overline{o})^2} \nonumber \\
 \frac{\mathrm{d}p_\mathrm{e}}{\mathrm{d}e} &=& \frac{g-\overline{o}}{(g+e-2\overline{o})^2}.
\label{eq:deriv}
\end{eqnarray}
\begin{figure}
\centerline{\includegraphics[width=8 cm]{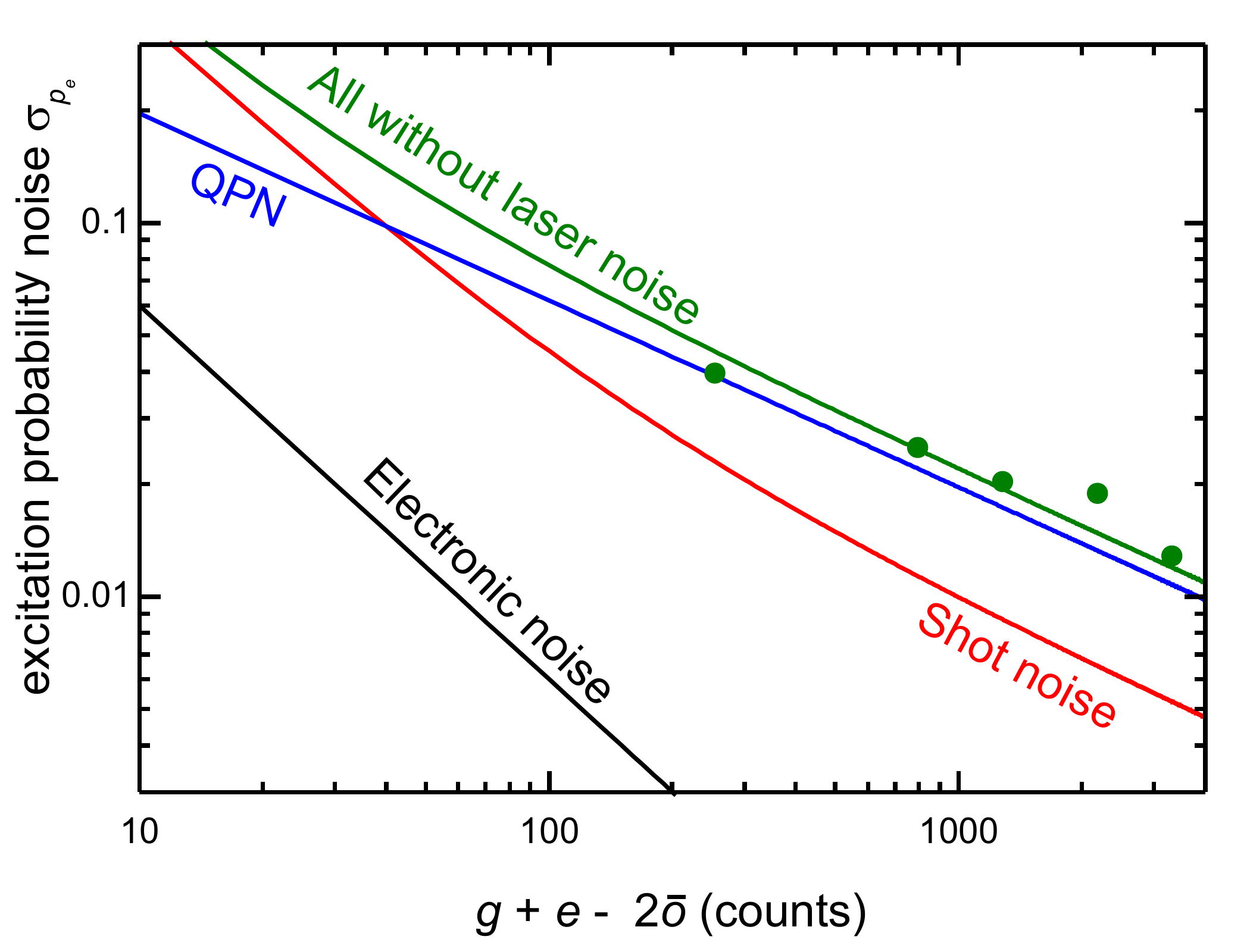}}
\caption{Experimental noise of the excitation probability with suppressed sensitivity to laser frequency noise ($\pi/2$-pulses, full circles) and estimated individual noise contributions. The QPN is based on a detection efficiency of 0.65~atoms/count. The shot noise calculation includes the noise of a typical background of 70~counts due to detection laser stray light. The green curve shows the summed noise according to Eq.~\ref{eq:noisemod}. The good agreement with the experimental data demonstrates the completeness of the model and the absence of laser noise in the data.}
\label{fig:noise}
\end{figure}
In Fig.~\ref{fig:noise}, we summarize the contributions to the detected excitation probability noise $\sigma_{p_\mathrm{e}}$ as a function of the total atom number as expressed by $g + e -2\overline{o}$. We see that the total noise agrees with the observations from Fig.~\ref{fig:QPN} very well, when we use a conversion factor of 0.65~atoms/count to calculate the QPN in Eq.~\ref{eq:QPN}. This is in reasonable agreement with an independent calibration derived from absorption measurements.

From this analysis we conclude that detection is limited by QPN for more than 200~counts, or 130~atoms. For 300~atoms this would lead to a frequency stability of our lattice clock of less than $6 \times 10^{-17}/\sqrt{\tau / \mathrm{s}}$, where we made use of Eqs.~\ref{eq:slope} and \ref{eq:QPN} and used a realistic interrogation time $T_\mathrm{\pi}=640$~ms with a cycle time of $T_\mathrm{c} = 1$~s \cite{fal14,hag14}. However, this estimate does not take into account the degradation of the stability due to the Dick effect, i.e.\ the aliasing of high frequency laser noise due to the non-continuous interrogation of the atoms.
%
%
%
%
%
%
\section{\label{sec:laser} Laser noise and Dick effect}
\begin{figure}
\centerline{\includegraphics[width=8 cm]{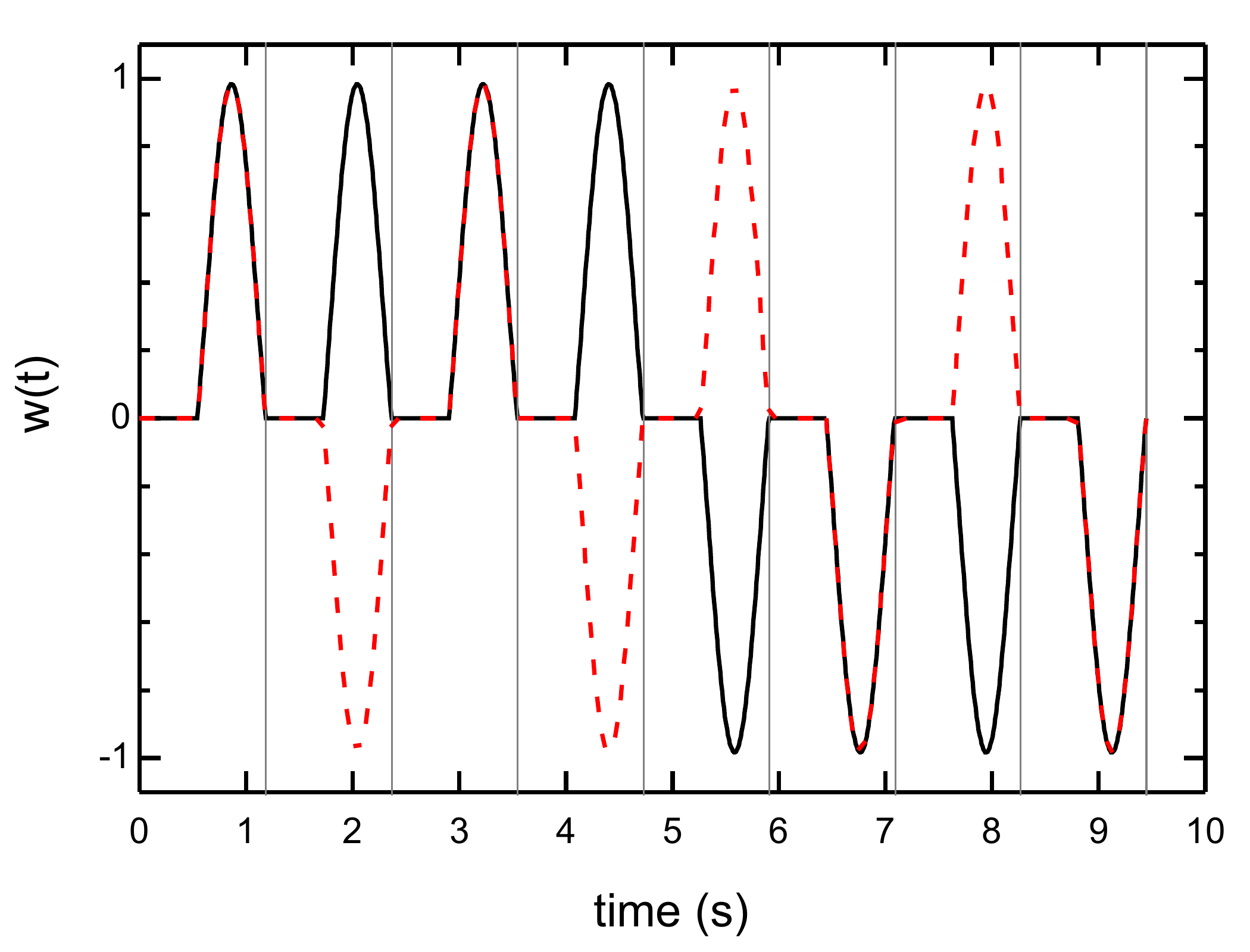}}
\caption{Sensitivity function $w(t)$ of interleaved stabilizations, shown for the cases of lumped (solid line) and distributed (dashed line) arrangements, as described in the text. Vertical grey lines indicate multiples of $T_\mathrm{c}$. The typical parameter values given in Sec.~\ref{sec:laser} were used. For a single stabilization, $w(t)$ consists only of the first four cycles shown for the lumped interleaved arrangement.} 
\label{fig:sensitivity}
\end{figure}
\begin{figure*}
\includegraphics[width=1.0\textwidth]{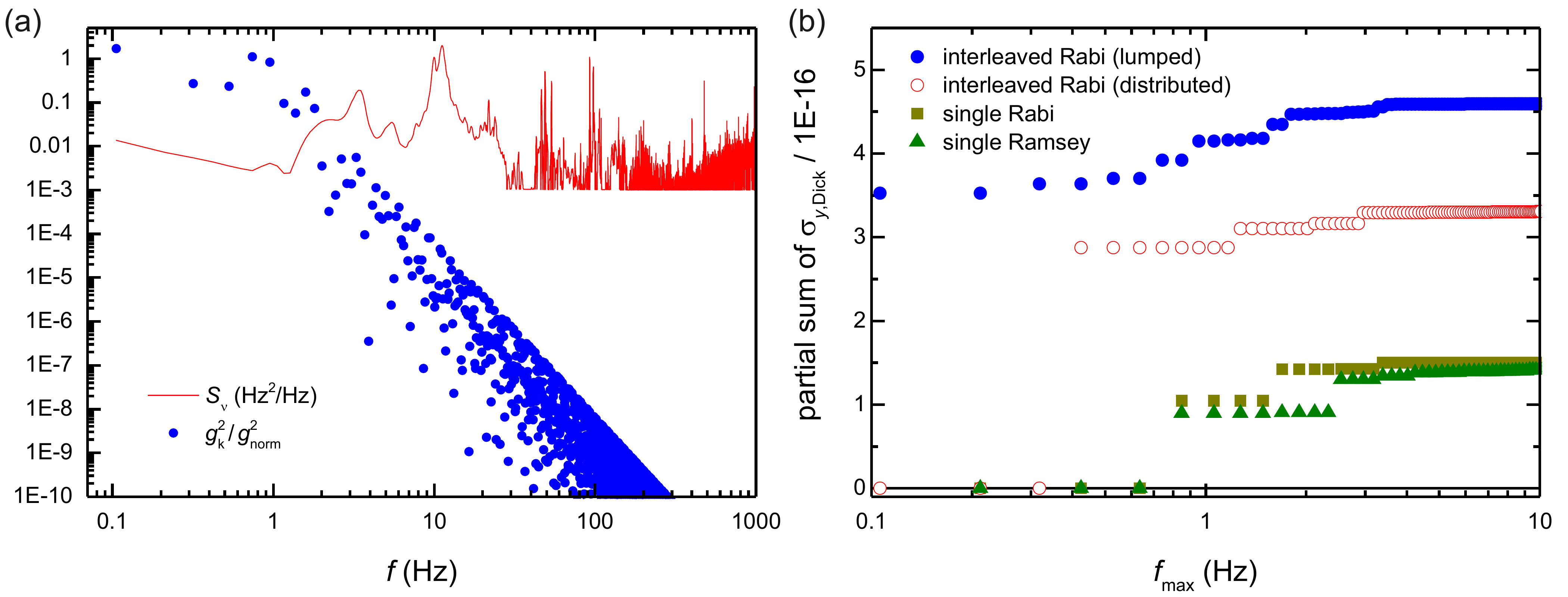}
\caption{
(a) Spectral power density of frequency fluctuations $S_\nu = \nu_0^2 S_y$ (solid line) and Fourier components $|g_k^2/g_\mathrm{norm}^2|$ (circles) for $k>0$ for interleaved stabilizations for parameter values as given in Sec.~\ref{sec:laser}. 
(b) Partial sums of $\sigma_{y,\mathrm{Dick}}^2(\tau)$ (Eq.~\ref{eq:dick}) as a function of the Fourier frequency $f_\mathrm{max} = k_\mathrm{max}/T$ at $\tau = 1$~s and for the parameter values given in Sec.~\ref{sec:laser}. Shown are the cases of interleaved stabilizations using either a lumped (solid circles) or a more stable distributed (open circles) probing sequence, as described in the text, and single stabilization using Rabi (solid squares) and alternatively Ramsey (solid triangles, same interrogation time with $\pi/2$-pulses of 100~ms length) interrogation.}
\label{fig:Dick}
\end{figure*}
After establishing the noise limitations on the optical clock instability from the atomic side, we will now show what frequency instability is introduced by the interrogation laser. Because of the non-continuous interrogation of the atoms, a stability degradation due to aliasing of noise -- the so-called Dick effect \cite{dic87} -- is expected. Its contribution, $\sigma_{y,\mathrm{Dick}}(\tau)$, to the fractional clock instability can be calculated by
\begin{equation}
\sigma_{y,\mathrm{Dick}}^2(\tau) = \frac{1}{\tau}\frac{1}{\left|g_0\right|^2} \sum^\infty_{k=1} S_y(k/T) \left| g_k \right|^2
\label{eq:dick}
\end{equation}
where $T= n T_\mathrm{c}$, $n$ is the number of interrogations in each stabilization cycle. The cycle duration $T_\mathrm{c}$ consists of the preparation time $T_\mathrm{D}$ and the interrogation time $T_\mathrm{\pi}$.

In Eq.~\ref{eq:dick}, the laser's single-sided power spectral density $S_y$ is evaluated at multiples of the inverse of the total duration $T$ of a complete stabilization cycle weighted with coefficients $g_k$, which are the Fourier components of a sensitivity function $w(t)$, and normalized by the dc Fourier coefficient $g_0$. The sensitivity function $w(t)$ describes the response of the detected excitation probability to interrogation laser frequency changes \cite{dic87} 
\begin{eqnarray}
\delta p_\mathrm{e} & = & \frac{1}{2}\int_0^T 2\pi\ \delta\nu\left(t\right) \cdot w\left(t\right) \mathrm{d}t \label{eq:sensFreq}\mbox{.}
\end{eqnarray}
The dc Fourier coefficient $g_0 = 1/\pi ~ \mathrm{d}p_\mathrm{e} /{\mathrm{d}\nu}$  in Eq.~\ref{eq:dick} relates the change in excitation probability to a constant change in laser frequency.

For a $\pi$-pulse of duration~$T_\mathrm{\pi}$ centered at $t=T_\mathrm{\pi}/2$ and a detuning $\Delta\approx\pm 0.40/T_\mathrm{\pi}$ of the interrogation laser, i.e., to the half-maximum point of the resonance, it can be found (equation 11 in \cite{dic87}) that
\begin{eqnarray}
w_\mathrm{single}(t)=\left\{
\begin{array}{ll}
\sin^2\vartheta \ \cos \vartheta \\ 
\times\left[(1-\cos\Omega_2)\sin\Omega_1 \right. \nonumber\\
\left. + (1-\cos\Omega_1)\sin\Omega_2\right] &   \mbox{during  pulse,}\nonumber\\
\\
0 & \mbox{elsewhere}
\end{array}
\right.
\label{eq:w}
\end{eqnarray}
with 
\begin{equation}
 \vartheta  = \frac{\pi}{2}-\arctan \left( 2T_\mathrm{\pi} \Delta \right) 
\end{equation}
and
\begin{eqnarray}
 \Omega_1  &=& \pi \sqrt{1+\left(2T_\mathrm{\pi} \Delta\right)^2}\times \frac{t}{T_\mathrm{\pi}} \nonumber \\
 \Omega_2  &=& \pi \sqrt{1+\left(2T_\mathrm{\pi} \Delta\right)^2} \times\frac{T_\mathrm{\pi}-t}{T_\mathrm{\pi}} \nonumber.
\end{eqnarray}

During clock operation, each stabilization cycle of the clock consists of four distinct interrogation (and preparation) sequences that interrogate the $m_F = \pm 9/2$ components on both slopes \cite{fal14}. Thus, a clock laser frequency correction is applied every $4T_\mathrm{c}$, and $w(t)$ is a fourfold series of $w_\mathrm{single}(t)$ (Fig.~\ref{fig:sensitivity}).

We use an interleaved stabilization scheme to evaluate most systematic shifts. The parameters of interest are alternated after each full clock stabilization cycle; separate digital servos correct the clock laser frequency independently for each configuration and generate a difference frequency signal. As the laser frequency noise is in general correlated between the interleaved interrogations, the Dick effect cannot be simply treated independently for each of the stabilizations. In analogy to the Dick effect for clock operation, a Dick effect can also be derived for the difference between the two interleaved stabilizations. The sensitivity function is composed of eight repetitions of $w_\mathrm{single}(t)$, where the sign is reversed between the first and last four (Fig.~\ref{fig:sensitivity}); a possible improvement to this lumped arrangement is discussed in Sec.~\ref{sec:stability} below. The conversion coefficient between the signal, i.e., the difference of excitation probabilities, and laser frequency changes is no longer given by $g_0$, but by the response to a unit frequency difference between both stabilization settings, which is
\begin{equation} 
g_\mathrm{norm} = \int_0^T \frac{1}{2} \left| w(t) \right| \mathrm{d}t. 
\end{equation}
Thus, the instability of the difference between the two interleaved stabilizations can be calculated from Eq. \ref{eq:dick}, with the above sensitivity function and replacing $g_0$ by $g_\mathrm{norm}$.

To calculate $\sigma_{y,\mathrm{Dick}}^2(\tau)$ from Eq.~\ref{eq:dick}, the laser noise spectrum is required. We have measured $S_y$ via three-cornered-hat comparisons with other lasers \cite{hae15a}. The noise spectrum is shown together with the $g_k$ for $k>0$ in Fig.~\ref{fig:Dick}a.

Having now all ingredients, the sum in Eq.~\ref{eq:dick} can be evaluated for $\tau = 1$~s and typical parameter values ($T_\mathrm{\pi} = 0.64$~s,  $T_\mathrm{D} = 0.54$~s, and  $T_\mathrm{c} = T_\mathrm{\pi} + T_\mathrm{D} = 1.18$~s). To visualize the contributions to $\sigma_{y,\mathrm{Dick}}$ from different Fourier frequencies, the partial sum of Eq.~\ref{eq:dick} up to the frequency $f_\mathrm{max} = k_\mathrm{max}/T$ is plotted in Fig.~\ref{fig:Dick}b. We see, that for a single stabilization, $\sigma_{y,\mathrm{Dick}}^2(\tau) $ is dominated by  $g_4$, whereas higher frequencies between 1~Hz to 10~Hz  contribute significantly for the interleaved stabilization. In fact, because of the symmetry of the single stabilization, all $g_k$ where $k$ is not a multiple of 4 are zero. Thus, the Dick effect is exactly the same as for a single interrogation cycle of length $T_\mathrm{c}$.
%
%
%
%
%
%
\section{\label{sec:stability}Lattice clock instability}
To estimate the instability of our lattice clock, the results from Secs.~\ref{sub:det} and \ref{sec:laser} were combined; the instabilities due to detection noise and the Dick effect have to be added in quadrature. For the difference of two interleaved stabilizations, the detection noise contribution is $\sqrt{2}$ times that of a single stabilization discussed in Sec.~\ref{sec:noise} due to the two independent stabilizations, whereas the Dick effect for the difference is accounted for by the combined sensitivity function discussed in Sec.~\ref{sec:laser}.

\begin{figure}
\centerline{\includegraphics[width=8 cm]{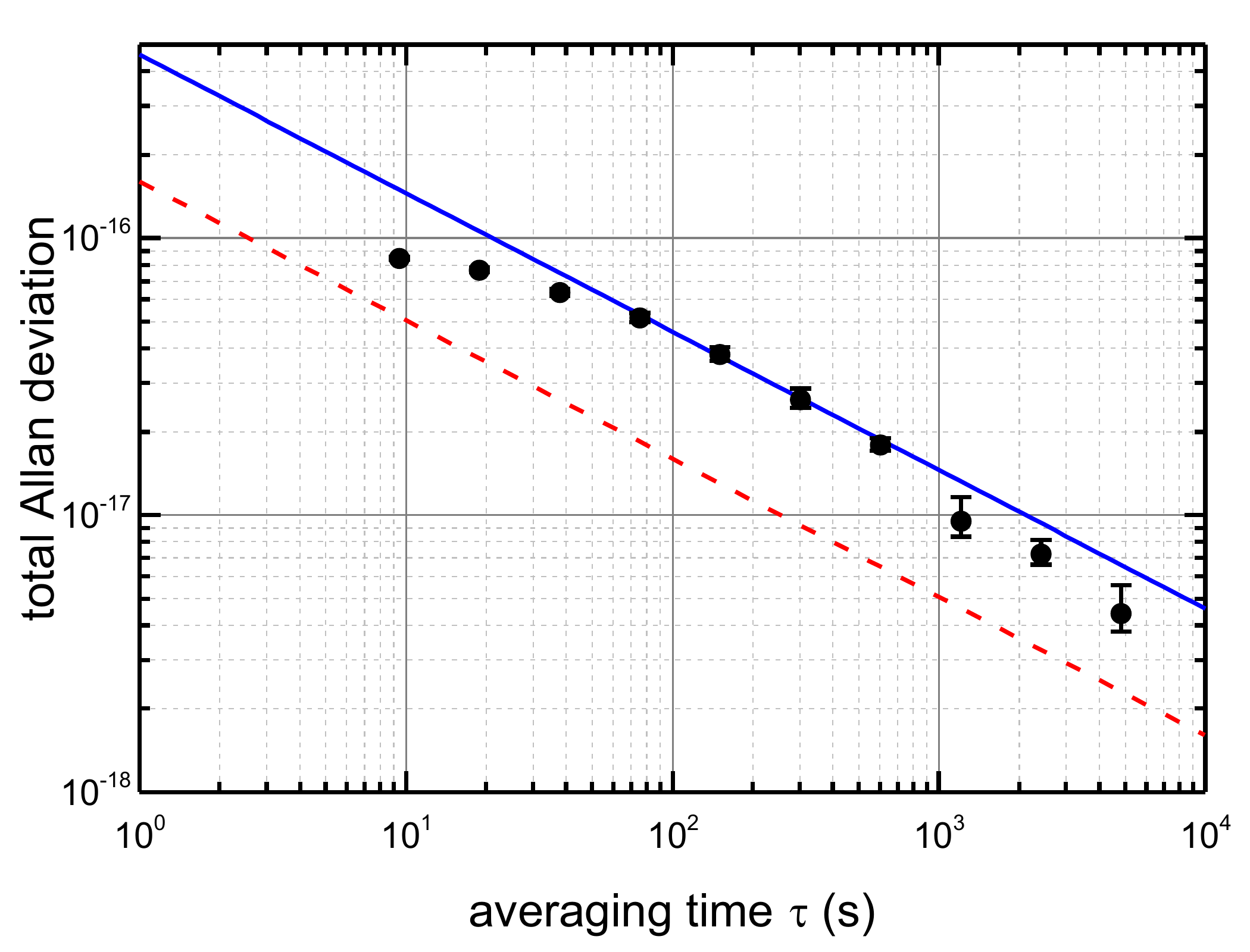}}
\caption{Allan deviation of the difference between two interleaved stabilizations with $T_\mathrm{\pi} = 0.64$~s,  $T_\mathrm{D} = 0.54$~s and $N \approx 360$ atoms (full circles). The fractional instabilities $\sigma_y(\tau)$ inferred from our analysis for interleaved stabilizations (solid blue line, $4.7 \times 10^{-16}/\sqrt{\tau/\mathrm{s}}$) and pure clock operation (dashed red line, $1.6 \times 10^{-16}/\sqrt{\tau/\mathrm{s}}$) are shown for comparison.}
\label{fig:AllanDev}
\end{figure}
From our analysis we infer a combined instability of $4.7 \times 10^{-16}/\sqrt{\tau/\mathrm{s}}$, with contributions of $4.6 \times 10^{-16}/\sqrt{\tau/\mathrm{s}}$ from the Dick effect and $6 \times 10^{-17}/\sqrt{\tau/\mathrm{s}}$ for each of the interleaved stabilizations from detection noise. Experimentally, an instability very close to that inferred from our analysis is found. Figure~\ref{fig:AllanDev} shows the Allan deviation of the difference signal observed in an interleaved stabilization for the experimental parameters given in Sec.~\ref{sec:laser} and $N \approx 360$ atoms. Thus, we conclude that the combined analysis of detection and laser noise constitutes a very good description of our optical lattice clock. 

The stability of interleaved stabilizations can be improved by modifying the sequence to use a distributed instead of a lumped arrangement, i.e.\ interleaving the interrogation sequences of each stabilization as well rather than arranging them in blocks of four, as shown in Fig.~\ref{fig:sensitivity}. We estimate that a combined instability of $3.4 \times 10^{-16}$ in 1~s can be achieved.

Based on the previous findings, we can infer the instability in clock operation with typical atom numbers of $N \approx 300$ atoms. Experimentally, we cannot determine this quantity directly, since we are lacking sufficiently stable oscillators to compare to. With this analysis we find $\sigma_y(\tau) = 1.6 \times 10^{-16}/\sqrt{\tau/\mathrm{s}}$ dominated by the Dick effect--induced instability of $\sigma_{y,\mathrm{Dick}}(\tau) = 1.5 \times 10^{-16}\sqrt{\tau/\mathrm{s}}$ (Fig.~\ref{fig:AllanDev}). A slight reduction of instability can be achieved by Ramsey interrogation instead of Rabi pulses (Fig.~\ref{fig:Dick}).

%
%
%
%
%
%
\section{\label{sec:concl}Conclusion}
We have given a full analysis of the detection noise of our Sr lattice clock. In combination with a Dick effect analysis of our new clock laser \cite{hae15a}, we are able to give a full description of our clock's observed frequency instability when differentially evaluating systematic frequency shifts. Since the agreement between model and observation is excellent, we are convinced that the instability of $\sigma_y(\tau) = 1.6 \times 10^{-16}/\sqrt{\tau/\mathrm{s}}$ we determine for normal clock operation is realistic. This is an exceptionally small instability, better than the so far published instabilities of optical clocks \cite{nic15,hin13,nic12}.

Although the residual noise of our clock laser \cite{hae15a} has been crucial for achieving this result, we also conclude that, even with one of the most advanced interrogation lasers available today and an efficient preparation scheme with a duty cycle of more than 50~\%, the clock instability is, already at very small atom numbers, limited by the Dick effect and thus the clock laser. This means that at present advanced squeezing or entanglement methods are not yet worthwhile.

First, the Dick effect--induced instability must be addressed. Ramsey interrogation offers a favorable sensitivity function as compared to a Rabi scheme, especially for high duty cycles \cite{wes10a}. Moreover, reducing the clock laser frequency noise between 1~Hz and 10~Hz, which is caused by seismic perturbations of the reference cavity, e.g.\ by active feed-forward \cite{lei11a}, could bring the clock instability to below $10^{-16}$ in one second. Finally, the duty cycle of clock laser interrogation could be increased: Nondestructive detection methods \cite{lod09}, which use, e.g., the phase-shift imprinted onto an off-resonant detection beam, allow keeping the majority of cold atoms in the lattice and thus reduce the required preparation time. Furthermore, few independent physics packages can use the same clock laser for interrogation to achieve a dead time--free observation of the laser frequency, thereby eliminating the Dick effect altogether. The detection noise presented in Sec.~\ref{sec:noise} would also be reduced by a factor of $\sqrt{2}$ owing to the dual interrogation per cycle. Such a setup leads to an ultimate instability of $4 \times 10^{-17}/\sqrt{\tau/\mathrm{s}}$ for the typical parameters discussed here and allows quantum projection noise--limited detection.

\begin{acknowledgments}
This work was performed within the framework of the Centre of Quantum Engineering and Space-Time Research (QUEST). Funding from the German Research Foundation DFG within the geo-Q collaborative research center CRC~1128 and research training group RTG~1729 is acknowledged, as well as funding within the ITOC and QESOCAS projects in the European Metrology Research Programme EMRP. The EMRP is jointly funded by the EMRP participating countries within EURAMET and the European Union. We thank Rodolphe Le Targat and S\'ebastien Bize from LNE-SYRTE, Observatoire de Paris for pointing out the importance of laser frequency correlations between the interleaved stabilizations.
\end{acknowledgments}


\begin{thebibliography}{22}
\expandafter\ifx\csname natexlab\endcsname\relax\def\natexlab#1{#1}\fi
\expandafter\ifx\csname bibnamefont\endcsname\relax
  \def\bibnamefont#1{#1}\fi
\expandafter\ifx\csname bibfnamefont\endcsname\relax
  \def\bibfnamefont#1{#1}\fi
\expandafter\ifx\csname citenamefont\endcsname\relax
  \def\citenamefont#1{#1}\fi
\expandafter\ifx\csname url\endcsname\relax
  \def\url#1{\texttt{#1}}\fi
\expandafter\ifx\csname urlprefix\endcsname\relax\def\urlprefix{URL }\fi
\providecommand{\bibinfo}[2]{#2}
\providecommand{\eprint}[2][]{\url{#2}}

\bibitem[{\citenamefont{Nicholson et~al.}(2015)\citenamefont{Nicholson,
  Campbell, Hutson, Marti, Bloom, McNally, Zhang, Barrett, Safronova, Strouse
  et~al.}}]{nic15}
\bibinfo{author}{\bibfnamefont{T.~L.} \bibnamefont{Nicholson}},
  \bibinfo{author}{\bibfnamefont{S.~L.} \bibnamefont{Campbell}},
  \bibinfo{author}{\bibfnamefont{R.~B.} \bibnamefont{Hutson}},
  \bibinfo{author}{\bibfnamefont{G.~E.} \bibnamefont{Marti}},
  \bibinfo{author}{\bibfnamefont{B.~J.} \bibnamefont{Bloom}},
  \bibinfo{author}{\bibfnamefont{R.~L.} \bibnamefont{McNally}},
  \bibinfo{author}{\bibfnamefont{W.}~\bibnamefont{Zhang}},
  \bibinfo{author}{\bibfnamefont{M.~D.} \bibnamefont{Barrett}},
  \bibinfo{author}{\bibfnamefont{M.~S.} \bibnamefont{Safronova}},
  \bibinfo{author}{\bibfnamefont{G.~F.} \bibnamefont{Strouse}},
  \bibnamefont{et~al.}, \bibinfo{journal}{Nature Com.}
  \textbf{\bibinfo{volume}{6}}, \bibinfo{pages}{6896} (\bibinfo{year}{2015}).

\bibitem[{\citenamefont{Hinkley et~al.}(2013)\citenamefont{Hinkley, Sherman,
  Phillips, Schioppo, Lemke, Beloy, Pizzocaro, Oates, and Ludlow}}]{hin13}
\bibinfo{author}{\bibfnamefont{N.}~\bibnamefont{Hinkley}},
  \bibinfo{author}{\bibfnamefont{J.~A.} \bibnamefont{Sherman}},
  \bibinfo{author}{\bibfnamefont{N.~B.} \bibnamefont{Phillips}},
  \bibinfo{author}{\bibfnamefont{M.}~\bibnamefont{Schioppo}},
  \bibinfo{author}{\bibfnamefont{N.~D.} \bibnamefont{Lemke}},
  \bibinfo{author}{\bibfnamefont{K.}~\bibnamefont{Beloy}},
  \bibinfo{author}{\bibfnamefont{M.}~\bibnamefont{Pizzocaro}},
  \bibinfo{author}{\bibfnamefont{C.~W.} \bibnamefont{Oates}}, \bibnamefont{and}
  \bibinfo{author}{\bibfnamefont{A.~D.} \bibnamefont{Ludlow}},
  \bibinfo{journal}{Science} \textbf{\bibinfo{volume}{341}},
  \bibinfo{pages}{1215} (\bibinfo{year}{2013}).

\bibitem[{\citenamefont{Nicholson et~al.}(2012)\citenamefont{Nicholson, Martin,
  Williams, Bloom, Bishof, Swallows, Campbell, and Ye}}]{nic12}
\bibinfo{author}{\bibfnamefont{T.}~\bibnamefont{Nicholson}},
  \bibinfo{author}{\bibfnamefont{M.}~\bibnamefont{Martin}},
  \bibinfo{author}{\bibfnamefont{J.}~\bibnamefont{Williams}},
  \bibinfo{author}{\bibfnamefont{B.}~\bibnamefont{Bloom}},
  \bibinfo{author}{\bibfnamefont{M.}~\bibnamefont{Bishof}},
  \bibinfo{author}{\bibfnamefont{M.}~\bibnamefont{Swallows}},
  \bibinfo{author}{\bibfnamefont{S.}~\bibnamefont{Campbell}}, \bibnamefont{and}
  \bibinfo{author}{\bibfnamefont{J.}~\bibnamefont{Ye}}, \bibinfo{journal}{Phys.
  Rev. Lett.} \textbf{\bibinfo{volume}{109}}, \bibinfo{pages}{230801}
  (\bibinfo{year}{2012}).

\bibitem[{\citenamefont{Jiang et~al.}(2011)\citenamefont{Jiang, Ludlow, Lemke,
  Fox, Sherman, Ma, and Oates}}]{jia11}
\bibinfo{author}{\bibfnamefont{Y.~Y.} \bibnamefont{Jiang}},
  \bibinfo{author}{\bibfnamefont{A.~D.} \bibnamefont{Ludlow}},
  \bibinfo{author}{\bibfnamefont{N.~D.} \bibnamefont{Lemke}},
  \bibinfo{author}{\bibfnamefont{R.~W.} \bibnamefont{Fox}},
  \bibinfo{author}{\bibfnamefont{J.~A.} \bibnamefont{Sherman}},
  \bibinfo{author}{\bibfnamefont{L.-S.} \bibnamefont{Ma}}, \bibnamefont{and}
  \bibinfo{author}{\bibfnamefont{C.~W.} \bibnamefont{Oates}},
  \bibinfo{journal}{Nature Photonics} \textbf{\bibinfo{volume}{5}},
  \bibinfo{pages}{158} (\bibinfo{year}{2011}).

\bibitem[{\citenamefont{Hagemann et~al.}(2013)\citenamefont{Hagemann, Grebing,
  Kessler, Falke, Lemke, Lisdat, Schnatz, Riehle, and Sterr}}]{hag13}
\bibinfo{author}{\bibfnamefont{C.}~\bibnamefont{Hagemann}},
  \bibinfo{author}{\bibfnamefont{C.}~\bibnamefont{Grebing}},
  \bibinfo{author}{\bibfnamefont{T.}~\bibnamefont{Kessler}},
  \bibinfo{author}{\bibfnamefont{S.}~\bibnamefont{Falke}},
  \bibinfo{author}{\bibfnamefont{N.}~\bibnamefont{Lemke}},
  \bibinfo{author}{\bibfnamefont{C.}~\bibnamefont{Lisdat}},
  \bibinfo{author}{\bibfnamefont{H.}~\bibnamefont{Schnatz}},
  \bibinfo{author}{\bibfnamefont{F.}~\bibnamefont{Riehle}}, \bibnamefont{and}
  \bibinfo{author}{\bibfnamefont{U.}~\bibnamefont{Sterr}},
  \bibinfo{journal}{IEEE Trans. Instrum. Meas.} \textbf{\bibinfo{volume}{62}},
  \bibinfo{pages}{1556} (\bibinfo{year}{2013}).

\bibitem[{\citenamefont{H{\"a}fner et~al.}(2015)\citenamefont{H{\"a}fner,
  Falke, Grebing, Vogt, Legero, Merimaa, Lisdat, and Sterr}}]{hae15a}
\bibinfo{author}{\bibfnamefont{S.}~\bibnamefont{H{\"a}fner}},
  \bibinfo{author}{\bibfnamefont{S.}~\bibnamefont{Falke}},
  \bibinfo{author}{\bibfnamefont{C.}~\bibnamefont{Grebing}},
  \bibinfo{author}{\bibfnamefont{S.}~\bibnamefont{Vogt}},
  \bibinfo{author}{\bibfnamefont{T.}~\bibnamefont{Legero}},
  \bibinfo{author}{\bibfnamefont{M.}~\bibnamefont{Merimaa}},
  \bibinfo{author}{\bibfnamefont{C.}~\bibnamefont{Lisdat}}, \bibnamefont{and}
  \bibinfo{author}{\bibfnamefont{U.}~\bibnamefont{Sterr}},
  \bibinfo{journal}{Opt. Lett.} \textbf{\bibinfo{volume}{40}},
  \bibinfo{pages}{2112} (\bibinfo{year}{2015}).

\bibitem[{\citenamefont{Derevianko and Pospelov}(2014)}]{der14}
\bibinfo{author}{\bibfnamefont{A.}~\bibnamefont{Derevianko}} \bibnamefont{and}
  \bibinfo{author}{\bibfnamefont{M.}~\bibnamefont{Pospelov}},
  \bibinfo{journal}{Nature Physics} \textbf{\bibinfo{volume}{10}},
  \bibinfo{pages}{933} (\bibinfo{year}{2014}).

\bibitem[{\citenamefont{Dick}(1988)}]{dic87}
\bibinfo{author}{\bibfnamefont{G.~J.} \bibnamefont{Dick}}, in
  \emph{\bibinfo{booktitle}{Proceedings of $19^{th}$ Annu. Precise Time and
  Time Interval Meeting, Redendo Beach, 1987}} (\bibinfo{publisher}{U.S. Naval
  Observatory}, \bibinfo{address}{Washington, DC}, \bibinfo{year}{1988}), pp.
  \bibinfo{pages}{133--147},
  \urlprefix\url{http://tycho.usno.navy.mil/ptti/1987/Vol%2019_13.pdf}.

\bibitem[{\citenamefont{Westergaard et~al.}(2010)\citenamefont{Westergaard,
  Lodewyck, and Lemonde}}]{wes10a}
\bibinfo{author}{\bibfnamefont{P.}~\bibnamefont{Westergaard}},
  \bibinfo{author}{\bibfnamefont{J.}~\bibnamefont{Lodewyck}}, \bibnamefont{and}
  \bibinfo{author}{\bibnamefont{Lemonde}}, \bibinfo{journal}{IEEE Trans.
  Ultrason. Ferroelectr. Freq. Control} \textbf{\bibinfo{volume}{57}},
  \bibinfo{pages}{623} (\bibinfo{year}{2010}).

\bibitem[{\citenamefont{Itano et~al.}(1993)\citenamefont{Itano, Bergquist,
  Bollinger, Gilligan, Heinzen, Moore, Raizen, and Wineland}}]{ita93}
\bibinfo{author}{\bibfnamefont{W.~M.} \bibnamefont{Itano}},
  \bibinfo{author}{\bibfnamefont{J.~C.} \bibnamefont{Bergquist}},
  \bibinfo{author}{\bibfnamefont{J.~J.} \bibnamefont{Bollinger}},
  \bibinfo{author}{\bibfnamefont{J.~M.} \bibnamefont{Gilligan}},
  \bibinfo{author}{\bibfnamefont{D.~J.} \bibnamefont{Heinzen}},
  \bibinfo{author}{\bibfnamefont{F.~L.} \bibnamefont{Moore}},
  \bibinfo{author}{\bibfnamefont{M.~G.} \bibnamefont{Raizen}},
  \bibnamefont{and} \bibinfo{author}{\bibfnamefont{D.~J.}
  \bibnamefont{Wineland}}, \bibinfo{journal}{Phys. Rev.~A}
  \textbf{\bibinfo{volume}{47}}, \bibinfo{pages}{3554} (\bibinfo{year}{1993}),
  \bibinfo{note}{see Also: Erratum Phys. Rev. A 51, 1717 (1995)}.

\bibitem[{\citenamefont{Rey et~al.}(2014)\citenamefont{Rey, Gorshkov, Kraus,
  Martin, Bishof, Swallows, Zhang, Benko, Ye, Lemke et~al.}}]{rey14}
\bibinfo{author}{\bibfnamefont{A.~M.} \bibnamefont{Rey}},
  \bibinfo{author}{\bibfnamefont{A.~V.} \bibnamefont{Gorshkov}},
  \bibinfo{author}{\bibfnamefont{C.~V.} \bibnamefont{Kraus}},
  \bibinfo{author}{\bibfnamefont{M.~J.} \bibnamefont{Martin}},
  \bibinfo{author}{\bibfnamefont{M.}~\bibnamefont{Bishof}},
  \bibinfo{author}{\bibfnamefont{M.~D.} \bibnamefont{Swallows}},
  \bibinfo{author}{\bibfnamefont{X.}~\bibnamefont{Zhang}},
  \bibinfo{author}{\bibfnamefont{C.}~\bibnamefont{Benko}},
  \bibinfo{author}{\bibfnamefont{J.}~\bibnamefont{Ye}},
  \bibinfo{author}{\bibfnamefont{N.~D.} \bibnamefont{Lemke}},
  \bibnamefont{et~al.}, \bibinfo{journal}{Annals of Physics}
  \textbf{\bibinfo{volume}{340}}, \bibinfo{pages}{311} (\bibinfo{year}{2014}).

\bibitem[{\citenamefont{Lemke et~al.}(2011)\citenamefont{Lemke, von Stecher,
  Sherman, Rey, Oates, and Ludlow}}]{lem11}
\bibinfo{author}{\bibfnamefont{N.~D.} \bibnamefont{Lemke}},
  \bibinfo{author}{\bibfnamefont{J.}~\bibnamefont{von Stecher}},
  \bibinfo{author}{\bibfnamefont{J.~A.} \bibnamefont{Sherman}},
  \bibinfo{author}{\bibfnamefont{A.~M.} \bibnamefont{Rey}},
  \bibinfo{author}{\bibfnamefont{C.~W.} \bibnamefont{Oates}}, \bibnamefont{and}
  \bibinfo{author}{\bibfnamefont{A.~D.} \bibnamefont{Ludlow}},
  \bibinfo{journal}{Phys. Rev. Lett.} \textbf{\bibinfo{volume}{107}},
  \bibinfo{pages}{103902} (\bibinfo{year}{2011}).

\bibitem[{\citenamefont{Ludlow et~al.}(2011)\citenamefont{Ludlow, Lemke,
  Sherman, Oates, Qu\'em\'ener, von Stecher, and Rey}}]{lud11}
\bibinfo{author}{\bibfnamefont{A.~D.} \bibnamefont{Ludlow}},
  \bibinfo{author}{\bibfnamefont{N.~D.} \bibnamefont{Lemke}},
  \bibinfo{author}{\bibfnamefont{J.~A.} \bibnamefont{Sherman}},
  \bibinfo{author}{\bibfnamefont{C.~W.} \bibnamefont{Oates}},
  \bibinfo{author}{\bibfnamefont{G.}~\bibnamefont{Qu\'em\'ener}},
  \bibinfo{author}{\bibfnamefont{J.}~\bibnamefont{von Stecher}},
  \bibnamefont{and} \bibinfo{author}{\bibfnamefont{A.~M.} \bibnamefont{Rey}},
  \bibinfo{journal}{Phys. Rev.~A} \textbf{\bibinfo{volume}{84}},
  \bibinfo{pages}{052724} (\bibinfo{year}{2011}).

\bibitem[{\citenamefont{Gibble}(2009)}]{gib09}
\bibinfo{author}{\bibfnamefont{K.}~\bibnamefont{Gibble}},
  \bibinfo{journal}{Phys. Rev. Lett.} \textbf{\bibinfo{volume}{103}},
  \bibinfo{pages}{113202} (\bibinfo{year}{2009}).

\bibitem[{\citenamefont{Wineland et~al.}(1992)\citenamefont{Wineland,
  Bollinger, Itano, Moore, and Heinzen}}]{win92}
\bibinfo{author}{\bibfnamefont{D.~J.} \bibnamefont{Wineland}},
  \bibinfo{author}{\bibfnamefont{J.~J.} \bibnamefont{Bollinger}},
  \bibinfo{author}{\bibfnamefont{W.~M.} \bibnamefont{Itano}},
  \bibinfo{author}{\bibfnamefont{F.~L.} \bibnamefont{Moore}}, \bibnamefont{and}
  \bibinfo{author}{\bibfnamefont{D.~J.} \bibnamefont{Heinzen}},
  \bibinfo{journal}{Phys. Rev.~A} \textbf{\bibinfo{volume}{46}},
  \bibinfo{pages}{R6797} (\bibinfo{year}{1992}).

\bibitem[{\citenamefont{Kessler et~al.}(2014)\citenamefont{Kessler, K\'om\'ar,
  Bishof, Jiang, S{\o}rensen, Ye, and Lukin}}]{kes14}
\bibinfo{author}{\bibfnamefont{E.~M.} \bibnamefont{Kessler}},
  \bibinfo{author}{\bibfnamefont{P.}~\bibnamefont{K\'om\'ar}},
  \bibinfo{author}{\bibfnamefont{M.}~\bibnamefont{Bishof}},
  \bibinfo{author}{\bibfnamefont{L.}~\bibnamefont{Jiang}},
  \bibinfo{author}{\bibfnamefont{A.~S.} \bibnamefont{S{\o}rensen}},
  \bibinfo{author}{\bibfnamefont{J.}~\bibnamefont{Ye}}, \bibnamefont{and}
  \bibinfo{author}{\bibfnamefont{M.~D.} \bibnamefont{Lukin}},
  \bibinfo{journal}{Phys. Rev. Lett.} \textbf{\bibinfo{volume}{112}},
  \bibinfo{pages}{190403} (\bibinfo{year}{2014}).

\bibitem[{\citenamefont{Falke et~al.}(2014)\citenamefont{Falke, Lemke, Grebing,
  Lipphardt, Weyers, Gerginov, Huntemann, Hagemann, Al-Masoudi, H{\"a}fner
  et~al.}}]{fal14}
\bibinfo{author}{\bibfnamefont{S.}~\bibnamefont{Falke}},
  \bibinfo{author}{\bibfnamefont{N.}~\bibnamefont{Lemke}},
  \bibinfo{author}{\bibfnamefont{C.}~\bibnamefont{Grebing}},
  \bibinfo{author}{\bibfnamefont{B.}~\bibnamefont{Lipphardt}},
  \bibinfo{author}{\bibfnamefont{S.}~\bibnamefont{Weyers}},
  \bibinfo{author}{\bibfnamefont{V.}~\bibnamefont{Gerginov}},
  \bibinfo{author}{\bibfnamefont{N.}~\bibnamefont{Huntemann}},
  \bibinfo{author}{\bibfnamefont{C.}~\bibnamefont{Hagemann}},
  \bibinfo{author}{\bibfnamefont{A.}~\bibnamefont{Al-Masoudi}},
  \bibinfo{author}{\bibfnamefont{S.}~\bibnamefont{H{\"a}fner}},
  \bibnamefont{et~al.}, \bibinfo{journal}{New J. Phys.}
  \textbf{\bibinfo{volume}{16}}, \bibinfo{pages}{073023}
  (\bibinfo{year}{2014}).

\bibitem[{\citenamefont{Falke et~al.}(2011)\citenamefont{Falke, Schnatz,
  Vellore~Winfred, Middelmann, Vogt, Weyers, Lipphardt, Grosche, Riehle, Sterr
  et~al.}}]{fal11}
\bibinfo{author}{\bibfnamefont{S.}~\bibnamefont{Falke}},
  \bibinfo{author}{\bibfnamefont{H.}~\bibnamefont{Schnatz}},
  \bibinfo{author}{\bibfnamefont{J.~S.~R.} \bibnamefont{Vellore~Winfred}},
  \bibinfo{author}{\bibfnamefont{T.}~\bibnamefont{Middelmann}},
  \bibinfo{author}{\bibfnamefont{S.}~\bibnamefont{Vogt}},
  \bibinfo{author}{\bibfnamefont{S.}~\bibnamefont{Weyers}},
  \bibinfo{author}{\bibfnamefont{B.}~\bibnamefont{Lipphardt}},
  \bibinfo{author}{\bibfnamefont{G.}~\bibnamefont{Grosche}},
  \bibinfo{author}{\bibfnamefont{F.}~\bibnamefont{Riehle}},
  \bibinfo{author}{\bibfnamefont{U.}~\bibnamefont{Sterr}},
  \bibnamefont{et~al.}, \bibinfo{journal}{Metrologia}
  \textbf{\bibinfo{volume}{48}}, \bibinfo{pages}{399} (\bibinfo{year}{2011}).

\bibitem[{\citenamefont{Lisdat et~al.}(2009)\citenamefont{Lisdat,
  Vellore~Winfred, Middelmann, Riehle, and Sterr}}]{lis09}
\bibinfo{author}{\bibfnamefont{C.}~\bibnamefont{Lisdat}},
  \bibinfo{author}{\bibfnamefont{J.~S.~R.} \bibnamefont{Vellore~Winfred}},
  \bibinfo{author}{\bibfnamefont{T.}~\bibnamefont{Middelmann}},
  \bibinfo{author}{\bibfnamefont{F.}~\bibnamefont{Riehle}}, \bibnamefont{and}
  \bibinfo{author}{\bibfnamefont{U.}~\bibnamefont{Sterr}},
  \bibinfo{journal}{Phys. Rev. Lett.} \textbf{\bibinfo{volume}{103}},
  \bibinfo{pages}{090801} (\bibinfo{year}{2009}).

\bibitem[{\citenamefont{Hagemann et~al.}(2014)\citenamefont{Hagemann, Grebing,
  Lisdat, Falke, Legero, Sterr, Riehle, Martin, and Ye}}]{hag14}
\bibinfo{author}{\bibfnamefont{C.}~\bibnamefont{Hagemann}},
  \bibinfo{author}{\bibfnamefont{C.}~\bibnamefont{Grebing}},
  \bibinfo{author}{\bibfnamefont{C.}~\bibnamefont{Lisdat}},
  \bibinfo{author}{\bibfnamefont{S.}~\bibnamefont{Falke}},
  \bibinfo{author}{\bibfnamefont{T.}~\bibnamefont{Legero}},
  \bibinfo{author}{\bibfnamefont{U.}~\bibnamefont{Sterr}},
  \bibinfo{author}{\bibfnamefont{F.}~\bibnamefont{Riehle}},
  \bibinfo{author}{\bibfnamefont{M.~J.} \bibnamefont{Martin}},
  \bibnamefont{and} \bibinfo{author}{\bibfnamefont{J.}~\bibnamefont{Ye}},
  \bibinfo{journal}{Opt. Lett.} \textbf{\bibinfo{volume}{39}},
  \bibinfo{pages}{5102} (\bibinfo{year}{2014}).

\bibitem[{\citenamefont{Leibrandt et~al.}(2011)\citenamefont{Leibrandt, Thorpe,
  and James C.~Bergquist}}]{lei11a}
\bibinfo{author}{\bibfnamefont{D.~R.} \bibnamefont{Leibrandt}},
  \bibinfo{author}{\bibfnamefont{M.~J.} \bibnamefont{Thorpe}},
  \bibnamefont{and} \bibinfo{author}{\bibfnamefont{T.~R.} \bibnamefont{James
  C.~Bergquist}}, \bibinfo{journal}{Opt. Express}
  \textbf{\bibinfo{volume}{19}}, \bibinfo{pages}{10278} (\bibinfo{year}{2011}).

\bibitem[{\citenamefont{Lodewyck et~al.}(2009)\citenamefont{Lodewyck,
  Westergaard, and Lemonde}}]{lod09}
\bibinfo{author}{\bibfnamefont{J.}~\bibnamefont{Lodewyck}},
  \bibinfo{author}{\bibfnamefont{P.~G.} \bibnamefont{Westergaard}},
  \bibnamefont{and} \bibinfo{author}{\bibfnamefont{P.}~\bibnamefont{Lemonde}},
  \bibinfo{journal}{Phys. Rev.~A} \textbf{\bibinfo{volume}{79}},
  \bibinfo{pages}{061401(R)} (\bibinfo{year}{2009}).

\end{thebibliography}

%
\end{document}